\documentclass[twocolumn,showpacs,preprintnumbers,amsmath,amssymb,prb,floatfix]{revtex4}

\usepackage{epsfig,psfrag}
\usepackage{dcolumn}
\usepackage{bm}
\usepackage{graphicx}
\usepackage{color}

\begin{document}

\title{Adiabatic polaron dynamics and Josephson effect in a
superconducting molecular quantum dot}

\author{Alex Zazunov and Reinhold Egger}

\affiliation{Institut f\"ur Theoretische Physik,
Heinrich-Heine-Universit\"at, D-40225 D\"usseldorf, Germany}

\date{\today}

\begin{abstract}
We study the Josephson current through a resonant level coupled to a
vibration mode (local Holstein model) in the adiabatic limit of low
oscillator frequency. A semiclassical theory is then appropriate and
allows us to consider the oscillator dynamics within the Born-Oppenheimer
approximation for arbitrary electron-vibration couplings.  The
resulting Fokker-Planck equation has been solved in the most relevant
underdamped limit and yields the oscillator distribution
function and the Josephson current. Remarkably,
a transition from single-well to double-well behavior of the
effective oscillator potential surface is possible and can be tuned by
variation of the superconducting phase difference.
The Josephson current is shown to be only weakly affected by the
electron-vibration coupling due to strong phonon localization near
the bottom of the potential surface.
\end{abstract}


\maketitle
\section{Introduction}

The field of molecular electronics continues to pose interesting scientific 
questions that are also of applied relevance.
Many aspects of charge transport through junctions containing a
single molecule have already been clarified,\cite{nazarov,nitzan} 
and relatively simple theoretical 
models\cite{boese,flensberg,cornaglia,mitra,paulsson,koch,martin,zazu2sq,zazu2,richter,egger,lothar,marten,pistolesi,fabrizio}
can often capture the essential physics in such devices;
see also Ref.~\onlinecite{nitzan2} for a recent review.
To quote just a few important experimental works, single-molecule transport
has been studied using different organic molecules,\cite{organic}
fullerenes,\cite{park,natelson,paul} carbon nanotubes,\cite{leroy,sapmaz}
and single hydrogen molecules between Pt leads.\cite{ruitenbeek}
When two \textit{superconducting} (instead of normal-state)
electrodes with a phase difference $\phi$ are attached
to the molecule, the Josephson effect \cite{golubov} implies that
an equilibrium current $I(\phi)$ can flow through the molecular junction.
The impressive experimental control over supercurrents through molecular
junctions achieved recently (see, for instance, Ref.~\onlinecite{chauvin}
and references therein) has been accompanied by first theoretical
studies investigating the effects of vibrational or conformational
molecular modes on the supercurrent.
In particular, the effect of just one harmonic vibration mode
coupled to a single-level quantum dot (``local Holstein model'')
has been considered in the superconducting version.
Analytical results have been obtained via perturbation theory
in the molecule-to-lead tunnel couplings\cite{novotny} or in
the electron-vibration coupling.\cite{zazu1,bcs-holstein1}
Other works have modelled the conformational mode as a two-level
system.\cite{zazu3} 

In this work, we consider the superconducting local Holstein model
describing a single spin-degenerate electronic state coupled to
the vibration mode and to two superconducting electrodes with
phase difference $\phi$.  We focus on the \textit{adiabatic regime}, where the
oscillator dynamics is slow on characteristic timescales of the
electronic motion,  and typically many oscillator quanta are excited under
 strong electron-vibration coupling.  As shown below, this situation
is analogous to a heavy Brownian particle in a fast non-Ohmic
fermionic environment.\cite{weiss} The oscillator distribution function and the
Josephson current can then be calculated by using a semiclassical
description of the oscillator dynamics.  Thereby, a nonperturbative
treatment of the electron-vibration coupling is possible and  controlled
calculations can be performed in so far unexplored parameter regimes.
Similar ideas have been employed before in the description
of nonequilibrium normal-state transport for this model,\cite{martin,pistolesi}
which are here generalized  to the superconducting case. 
We  address the equilibrium case (no bias voltage), where the 
phase difference $\phi$ is the
relevant control parameter coupling to the oscillator's motion.
Note that typical superconducting gap scales are of the order of 
$\Delta\approx 1$~meV, and in many experimentally studied
cases,\cite{weig,garcia,huettel,adrian} the relevant vibrational energy scale is
significantly smaller than $\Delta$ and our theory is directly applicable.

The structure of this paper is as follows. In Sec.~\ref{sec2},
we discuss the model and introduce our semiclassical approach.
In Sec.~\ref{sec3}, we then consider the oscillator dynamics
within the Born-Oppenheimer approximation, and we derive
the Fokker-Planck equation for the distribution function in
energy space.  The approach is employed to obtain the results
presented in Sec.~\ref{sec4}, followed by
a discussion and some conclusions in Sec.~\ref{conc}.
We mostly use units where $e=\hbar=k_{B}=1$.

\section{Model and semiclassical approach} \label{sec2}

\subsection{Model}

We start by describing a minimal model of a molecular quantum dot sandwiched by
two superconducting leads.  Similar to the normal-state case,\cite{nitzan2}
this model can capture essential aspects of the relevant physics in such
devices.  Writing the full Hamiltonian
$H = H_0 + H_T + H_{\rm leads}$, the term $H_0$ describes the isolated ``molecule'',
including the vibration mode and its coupling to the electronic state.
$H_T$ refers to the electronic tunneling Hamiltonian connecting the molecular
level to the superconducting electrodes, and $H_{\rm leads}$ describes
the $s$-wave BCS superconducting leads with phase difference $\phi$.
In this work, we only discuss the equilibrium case where both
leads have the same chemical potential.  Concerning $H_0$, we assume that only one 
spin-degenerate molecular electronic state is relevant, with (bare) energy $\epsilon_0$.
The corresponding fermion operator is $d_\sigma$ for spin projection
$\sigma=\uparrow,\downarrow$.  The molecular dot is supposed to also host 
a dominant vibration mode of frequency $\Omega$, and we retain only this 
quantum oscillator mode with dimensionless canonically conjugate operators $x$ and $p$. 
Following standard arguments,\cite{nitzan2} usually the most important coupling ($\lambda$)
to $d_\sigma$ is contained in a dot Hamiltonian of the form
\begin{equation} \label{H0}
H_0 = ( \epsilon_0 + \lambda x ) \sum_\sigma
\left( d^\dagger_\sigma d_\sigma - \frac12 \right) +
\frac{\Omega}{2}  ( p^2 +  x^2 ) .
\end{equation}
In this representation, the oscillator acts as a charge parity detector,
since it is displaced by the dot charge $\hat{n} = \sum_\sigma d^\dagger_\sigma d^{}_\sigma$
only for even $n = \{0,2\}$.  It is convenient to employ the Nambu formalism, where
fermion operators are combined in the Nambu spinors $d = ( d^{}_\uparrow , 
d^\dagger_\downarrow)^T$ and $\psi^{}_{jk} =(\psi_{jk, \uparrow},
\psi^\dagger_{j(-k), \downarrow})^T$ for electrons in the left or right
lead ($j=L/R$) with momentum $k$.
The leads are then described by a pair of BCS Hamiltonians,
\begin{equation}
H_{\rm leads} = \sum_{j,k}
\psi^\dagger_{jk} \left( \epsilon_k
\sigma_z + \Delta \sigma_x \right) \psi_{jk} ,
\end{equation}
with normal-state dispersion $\epsilon_k$ and BCS gap $\Delta$ (taken real
and positive); for simplicity, we consider identical superconductors.
The standard Pauli matrices $\sigma_{x,y,z}$ and $\sigma_0={\rm diag}(1,1)$
act in Nambu space.  Tunneling of electrons between the dot and 
the leads corresponds to
\begin{equation} \label{HT}
H_T = \sum_{j=L,R=+,-} \sum_{k} t_0
\psi^\dagger_{jk}  \, \sigma_z  e^{\pm i \sigma_z \phi/4} \, d + {\rm h.c.},
\end{equation}
where we assume that both dot-to-lead tunnel couplings $t_0$ are equal and $k$-independent;
the generalization to asymmetric cases is straightforward.
The superconducting phase difference $\phi$ enters via the phase factor dressing the
tunnel matrix element. Finally, we define the hybridization energy 
$\Gamma= \pi \nu_0 |t_0|^2$, where $\nu_0 = 2 \sum_k \delta (\epsilon_k)$
is the normal density of states in the leads.

We next employ the real-time path integral technique\cite{nazarov,weiss} to derive an
effective action for the oscillator alone, i.e., we
integrate out all electronic degrees of freedom.
 Although we study an equilibrium problem, it is technically easier to obtain
the semiclassical limit from the real-time Keldysh technique.\cite{nazarov}
We thus introduce the standard forward [backward] branch
of the Keldysh contour, with oscillator trajectories $x_1(t)$  $[x_2(t)]$.
These define the \textit{classical} trajectory $x(t) = ( x_1 + x_2) /2$
and the \textit{quantum} part $y(t) = x_1 - x_2$.
The path-integral expression for the time evolution operator of the
system then takes the form
\begin{equation} \label{calZ}
{\cal Z} = \int {\cal D} x  {\cal D} y \, e^{i (S_0 + S_e )} ,
\end{equation}
where  the action of the uncoupled oscillator is
\begin{equation}
S_0 = - \int dt  \ y ( \Omega^{-1} \ddot{x} + \Omega x ) ,
\end{equation}
and $S_e$ is an influence functional which results from
tracing out all fermionic variables,\cite{nazarov,weiss}
\begin{eqnarray} \label{Sbath}
S_e & = & -i \, {\rm Tr} \ln \left( \check G^{-1} - \frac{\lambda}{2}
\sigma_z y \right) \\ \nonumber &=&
-i \, {\rm Tr} \ln \check G^{-1} +
i \sum_{n = 1}^\infty \frac{( \lambda / 2)^n}{n} \, {\rm Tr}
\left( \check G \sigma_z y \right)^n .
\end{eqnarray}
The trace operation ``Tr'' extends over Nambu, Keldysh, and time (or energy)
space, while the symbols ``Tr$_N$'' (``Tr$_K$'') used below will refer to a trace over Nambu
(Keldysh) space only.
$\check G$ denotes the Keldysh Green's function (GF) of the dot for
given classical trajectory ($y=0$); the check notation $(\check{~})$ indicates
the $2\times 2$ Keldysh structure.  In terms of the Nambu spinors $d$,
$\check G(t_1,t_2) = -i \langle {\cal T}_C
 [ d(t_1) d^\dagger(t_2) ] \rangle$,
where ${\cal T}_C$ is the time-ordering operator along the Keldysh contour.
It is convenient to express the GF $\check G(t_1,t_2)\equiv \check G(t;\tau)$
with $t=(t_1+t_2)/2$ and $\tau = t_1 - t_2$ in the Wigner (``mixed'') representation.
We will also frequently employ the Fourier transformed expression, 
$\check{G}(t; \tau) = (2\pi)^{-1}\int d \omega \, e^{- i \omega \tau} \,
 \check{G}(t; \omega).$

According to Eq.~(\ref{H0}), for a given classical
trajectory $\{ x(t) \}$ of the oscillator,
the dot Keldysh GF can be obtained from the Dyson equation
\begin{equation}\label{GGG}
\check G^{-1} = \check G_0^{-1} - \lambda x(t) \sigma_z \check \tau_z  ,
\end{equation}
which formally represents an infinite-dimensional matrix 
equation in time (or energy) space and in Nambu-Keldysh space.  
The (inverse) GF in the absence of the electron-vibration coupling is 
\begin{equation} \label{GG0}
\check G _0^{-1} = \left( i \partial_t - \epsilon_0 \sigma_z \right)
 \check \tau_z - \check \Sigma ,
\end{equation}
where the Pauli matrices $\check \tau_{x,y,z}$ act in Keldysh space.
The self-energy $\check\Sigma$ originates from the integration
over the lead fermions. In frequency representation, the retarded and advanced 
components (in Nambu space) are given by\cite{nazarov,golubov}
\begin{equation}
\Sigma^{R/A}(\omega) = -i \Gamma \frac{ \omega \sigma_0 -
 \Delta \cos(\phi/2) \sigma_x }{\sqrt{(\omega\pm i0^+)^2-\Delta^2}}  ,
\end{equation}
while the Keldysh component follows from the standard equilibrium relation,
\begin{equation}\label{sk}
\Sigma^K(\omega) = f(\omega) \left[ \Sigma^R(\omega)-
\Sigma^A(\omega)\right],\quad f(\omega)=\tanh(\omega/2T).
\end{equation}
These components determine the Keldysh matrix structure according to
($\nu=\pm$ denotes the upper/lower branch of the Keldysh contour)
\begin{equation}
\check \Sigma_{\nu\nu'}(\tau) = \frac12 \int \frac{d\omega}{2\pi}
e^{-i\omega\tau} \left[ \nu \Sigma^R + \nu' \Sigma^A + \nu\nu' \Sigma^K\right](\omega).
\end{equation}
Similarly, the Keldysh GF $\check G$ can be decomposed into the
retarded [$G^R$], advanced [$G^A$], and Keldysh [$G^K$] components.
Note that so far our expressions are exact.

\subsection{Semiclassical approach}

In this paper, our main interest concerns the adiabatic case of a
slow oscillator, where $\Omega$ is the smallest physical frequency scale.
In this limit, the kinetic term in $S_0$ favors small quantum fluctuations $y(t)$,
and a semiclassical approach expanding in $\{y(t)\}$ becomes possible.
(This approximation can also be justified in the limit of high temperatures.)
For the normal case ($\Delta=0$), such an approach has been worked out in detail
for nonequilibrium transport in Refs.~\onlinecite{martin} and \onlinecite{pistolesi}.
It constitutes a controlled appoximation for $\Omega\ll \Gamma$ and arbitrary $\lambda$.
In the superconducting case, we instead require $\Omega\ll {\rm min}(\Delta,\Gamma)$.

Within a semiclassical approach, we thus evaluate the action $S_e$ up to
second order in the quantum amplitude $y(t)$ while keeping the full nonlinear
dependence on the classical trajectory $x(t)$,
\begin{equation}
S_e = -i \, {\rm Tr} \ln \check G^{-1}
+ S_e^{(1)} + S_e^{(2)} + {\cal O}(y^3) .
\end{equation}
The first-order term is $S_e^{(1)} = \int dt \, {\cal F}(t)  y(t)$,
with the total \textit{force} exerted by electrons on the oscillator
\begin{equation}
{\cal F}(t) = \frac{i \lambda}{2} {\rm Tr}_N \left( G^K(t,t) \sigma_z \right) =
\lambda \Big( 1 - \langle \hat{n}(t) \rangle \Big).
\end{equation}
Using the Dyson equation (\ref{GG0}), some algebra yields
\begin{equation} \label{Sbath1v2}
{\cal F}(t) = F_e(t) - \int^t dt' \eta(t,t') \dot{x}(t') ,
\end{equation}
with the time-local part of the force,
\begin{equation} \label{Fbath}
F_e(t) = \frac{i\lambda}{2} \, {\rm Tr}_N \left[ G^K_0(t,t) \sigma_z \right]  
+ \eta(t,t) x(t) .
\end{equation}
Here the equal-time value of the damping kernel is
\begin{equation}\label{etatt}
\eta(t,t) = \frac{i \lambda^2}{2}
\int d t' \, {\rm Tr}_{N,K} \left( \check{G}_0(t-t') \check \tau_z \sigma_z
\check{G}(t', t) \sigma_z \right).
\end{equation}
The second term in Eq.~(\ref{Sbath1v2}) describes retarded damping, where
the Wigner representation of the real-valued damping kernel $\eta(t,t')$ 
with $t>t'$ is obtained by solving the equation
\begin{eqnarray} \label{SAprr:eta}
&& \left(\frac12 \partial_t + i \omega \right) \eta(t; \omega)
= \frac{\lambda^2}{4} \sum_{s = \pm} s \int \frac{d \omega'}{ 2 \pi }
\\ \nonumber && \times {\rm Tr}_N \Big[ A_0(t; \omega' + s \omega/2)
\sigma_z G^K(t; \omega' - s \omega/2) \sigma_z \\ \nonumber && -
G^K_0(t; \omega' + s \omega/2) \sigma_z A(t; \omega' -
s \omega/2) \sigma_z \Bigr ] .
\end{eqnarray}
Here $A = i  (G^R - G^A )$ is the full spectral function of the dot
(including the electron-vibration coupling), while $A_0$ refers to the
corresponding $\lambda=0$ case.   
The second-order \textit{noise term} is from Eq.~(\ref{Sbath}) given by
\begin{equation} \label{Sbath2}
S_e^{(2)} = \frac{i}{2} \int dt dt' \, y(t) K(t,t') y(t'),
\end{equation}
where the fluctuation kernel has the Wigner representation 
\begin{eqnarray} \nonumber
K(t; \omega)&=& \frac{\lambda^2}{4}  \int \frac{d \omega'}{2 \pi}
{\rm Tr}_N \Bigl( [A+iG^K](t; \omega' + \omega/2) \\ \label{SAppr:K}
&\times& \sigma_z [A-iG^K](t; \omega' - \omega/2) \sigma_z \Bigr) .
\end{eqnarray}

\subsection{Weak coupling limit: Fluctuation-dissipation theorem}

So far no approximations have been made in treating $S_e^{(1,2)}$, and
the above expressions are exact.  Before we address the adiabatic
regime of small $\Omega$,
it is instructive to briefly consider the case of small $\lambda$ but
arbitrary $\Omega$.  In that case, the full GF $\check{G}$ entering
Eqs.~(\ref{etatt}), (\ref{SAprr:eta}) and (\ref{SAppr:K})
can be replaced by the free GF $\check{G}_0$. Taking into account that
$G_0^K(\omega) = f(\omega) \left[ G_0^R(\omega) - G_0^A(\omega)
\right]$, cf.~Eq.~(\ref{sk}), we find from Eq.~(\ref{SAprr:eta})
\begin{eqnarray} \label{etawigner}
\eta(\omega) &=& \frac{\lambda^2}{ 2 \omega} \int \frac{d \omega'}{2 \pi} \,
\left[ f(\omega' + \omega/2)- f(\omega' - \omega/2) \right]\\ \nonumber
&\times& {\rm Tr}_N \left[ A_0(\omega' + \omega/2) \sigma_z
A_0(\omega' - \omega/2) \sigma_z \right].
\end{eqnarray}
Some algebra shows that the fluctuation kernel $K (\omega)$
in Eq.~(\ref{SAppr:K}) can also be expressed in terms of $\eta(\omega)$,
\begin{equation} \label{weakcoupl:FDT}
K(\omega) = \omega \left[n_B(\omega) + 1 \right ] \eta(\omega) ,
\end{equation}
where $n_B(\omega) = (e^{\omega/T} - 1)^{-1}$ is the Bose-Einstein
function.  

Equation (\ref{weakcoupl:FDT}) constitutes the
well-known fluctuation-dissipation theorem\cite{weiss} for weak
electron-vibration coupling and provides a consistency check for
our formalism.  In the normal state ($\Delta=0$),
the damping kernel $\eta(\omega)$ is often assumed
to be a smooth function of $\omega$, which is then approximated by a constant,
$\eta_0=\eta (\omega= 0)$, according to Eq.~(\ref{etawigner}).
Under this approximation, the damping constant entering the equation of
motion is just $\eta_0/2$, as follows from the resulting
first-order action, $S_e^{(1)} = \int dt \, y [ F_e -  (\eta_0/2) \dot{x} ]$.
In the high-temperature limit, the fluctuation kernel
then describes white noise, $K(\tau) =  \eta_0 T \delta(\tau)$.
In the superconducting case ($\Delta \neq 0$), however,
the above kernels may not exhibit the assumed spectral smoothness.
The presence of Andreev bound states is known to cause
singular behavior of the dot spectral function $A(t; \omega)$
in the subgap region $ |\omega|<\Delta$.
We will therefore take into account the electron
damping and fluctuation effects on the oscillator's motion
throughout the whole spectral range.
Furthermore, we now go beyond the weak-coupling limit and
consider a nonperturbative theory in the electron-vibration coupling $\lambda$.

\section{Adiabatic regime}
\label{sec3}

\subsection{Born-Oppenheimer approximation}

Next we turn to the oscillator dynamics in the adiabatic regime
realized for $\Omega \ll {\rm min}(\Gamma, E_a(\phi))$,
where $E_a(\phi)$ is the Andreev level energy (see below).
In particular, we require $\Omega\ll \Delta$, which also
implies that normal-state results do not follow
from the expressions below by sending $\Delta\to 0$.
Since the oscillator dynamics is now much slower than
the electronic motion, we can invoke the Born-Oppenheimer
(BO) approximation.\cite{weiss}
The dot GF $\check G$ is thereby approximated by
the \textit{adiabatic Green's function} $\check {\cal G}(t;\omega)$
whose inverse for given $t,\omega$ follows from the Dyson equation
\begin{equation} \label{calG}
\check{{\cal G}}^{-1}(t;\omega) = \check{G}_0^{-1}(\omega) -
\lambda \sigma_z \check{\tau}_z x(t) ,
\end{equation}
which now is a simple $4\times 4$ matrix (in Keldysh-Nambu space)
relation.  [We often denote $\check{{\cal G}}(x(t);\omega) =
\check{{\cal G}}(t;\omega)$.] This GF describes a noninteracting dot
whose time-dependent energy level $\epsilon (t) = \epsilon_0 + \lambda x(t)$
is determined by the instantaneous displacement $x=x(t)$ of the oscillator.
Equation (\ref{calG}) is formally obtained from the Dyson equation
(\ref{GGG}) for $\check{G}(t_1,t_2)$ in the mixed representation (with
$t_{1,2} = t \pm \tau/2$),
\begin{eqnarray*}
\check{G}(t;\tau) &=& \check{G}_0(\tau) + \lambda \int dt'
\check{G}_0(t_1-t') \check{\tau}_z \sigma_z \\
&\times& \left ( x(t) + ( t' - t ) \dot{x}(t) + \ldots \right) \\
&\times& \left( 1 -  \frac{t_1 - t'}{2} \partial_t + \ldots \right)
\check{G}(t;t'-t_2).
\end{eqnarray*}
Noting that $\partial_t$ corresponds to $\dot x \partial_x$,
all derivative terms are of order ${\cal O}(\dot x)$
and should therefore be neglected within the BO approximation.
Using the analogy to an effectively noninteracting quantum dot level,
the retarded component of the adiabatic GF follows in the form\cite{zazu1}
\begin{equation} \label{calGRomega}
{\cal G}^R(x;\omega) = \frac{\omega (1+\alpha_\omega) + \epsilon(x) \sigma_z +
\alpha_\omega \Delta \cos(\phi/2) \sigma_x} {{\cal D}(x; \omega)},
\end{equation}
where we introduce the auxiliary quantity
\begin{equation}\label{alphadef}
\alpha_\omega= \frac{i\Gamma}{\sqrt{(\omega+i0^+)^2-\Delta^2}}
\end{equation}
and the denominator is given by
\begin{equation} \label{calD}
{\cal D}(x; \omega) = \omega^2 (1+\alpha_\omega)^2- \epsilon^2(x) -
\alpha^2_\omega \Delta^2 \cos^2(\phi/2).
\end{equation}
The resulting adiabatic spectral function ${\cal A}(x; \omega)=
{\cal A}_a+{\cal A}_c$ receives contributions from
Andreev levels (${\cal A}_a$, for $|\omega| < \Delta$)
and from quasiparticle continuum states (${\cal A}_c,$ for $|\omega|>\Delta$).
The Andreev level spectral function is given by
\begin{eqnarray}
{\cal A}_a(x; \omega) &=& \frac{2 \pi}{\partial_\omega {\cal D}(x; \omega)}
\delta \left( |\omega| - E_a(x) \right) \\
\nonumber &\times&
\left[ \omega(1+\alpha_\omega) + \epsilon(x) \sigma_z +
\alpha_\omega \Delta \cos(\phi/2) \sigma_x \right],
\end{eqnarray}
where the $\phi$-dependent Andreev level energy $E_a(x) \in [0, \Delta)$
is a non-negative root of the equation ${\cal D}(x; E_a) = 0$.
The other components of the adiabatic GF then follow from ${\cal G}^A=
[{\cal G}^R]^\dagger$ and the equilibrium relation
\begin{equation}\label{calgk}
{\cal G}^K(t;\omega) = - i f(\omega) {\cal A}(x(t);\omega).
\end{equation}

Next we address the corresponding adiabatic expressions for
the damping and fluctuation kernels.
Since the damping kernel in Eq.~(\ref{Sbath1v2}) is multiplied
by $\dot x$, it is sufficient to replace $\check G \to {\cal G}$
in the calculation of the damping kernel $\eta$.  Using Eqs.~(\ref{etatt})
and (\ref{calG}), the time-local force (\ref{Fbath}) reads
\begin{equation} \label{Fbath2}
F_e(x) = \frac {\lambda}{2} \int \frac{d \omega}{ 2 \pi}\, f(\omega) \,
{\rm Tr}_N \left( \sigma_z {\cal A}(x;\omega) \right) ,
\end{equation}
while the damping [Eq.~(\ref{SAprr:eta})] and 
fluctuation [Eq.~(\ref{SAppr:K})] kernels are
\begin{eqnarray} \label{BO:etaK}
\eta(t;\omega) &=& \tilde \eta (x(t),0; \omega), \\ \nonumber
K(t;\omega) &=& \omega [n_B(\omega) + 1 ]\, \tilde{\eta}(x(t),x(t);\omega),
\end{eqnarray}
with a generalized ``dissipation function''
\begin{eqnarray}\nonumber 
&&\tilde \eta(x,x'; \omega) = \frac{\lambda^2}{4\omega}
\int \frac{d \omega'}{2 \pi } [ f(\omega' + \omega/2) -
f(\omega' - \omega/2)] \\ \nonumber &&\times
\sum_{s = \pm}\, {\rm Tr}_N \left( {\cal A}(x; \omega' + s \omega/2) \sigma_z
{\cal A}(x'; \omega' - s \omega/2) \sigma_z  \right) \\
&& \quad =\tilde\eta(x,x';-\omega)=\tilde \eta(x',x;\omega).
\label{BO:tildeEta}
\end{eqnarray}
For small $\lambda$, the $x$-dependence in Eq.~(\ref{BO:etaK})
can be neglected, and we recover the
fluctuation-dissipation theorem (\ref{weakcoupl:FDT}).
The decomposition ${\cal A}={\cal A}_a+{\cal A}_c$
implies that $\eta=\eta_a+\eta_c$ and $K=K_a+K_c$
separate into contributions from the Andreev level states and from
continuum states.  Mixed terms involving transitions between Andreev level and
continuum states turn out to be always strongly suppressed in the
adiabatic regime due to the presence of an energy threshold $\Delta - E_a$
in the fermionic spectrum.  As a result, such terms can safely be neglected.

At this point, we pause and summarize what we have achieved so far.
The total effective action of the oscillator within the BO approximation is
\begin{eqnarray}
S & = & - \int dt \, y \left[ \Omega^{-1} \ddot{x} - F(x) +
\int^t dt' \, \eta(t,t') \, \dot{x}(t') \right]  \nonumber  \\
\label{Seff} &+&\frac{i}{2} \int dt dt' \, y(t) K(t,t') y(t') -i \,
{\rm Tr} \ln \check{\cal G}^{-1} ,
\end{eqnarray}
where $F(x) = - \Omega x + F_e(x)$ is the total potential
force. 
The Wigner representation of the
dissipation and fluctuation kernels is
\begin{eqnarray} \label{etaX}
\eta(t; \tau) &=& \int_0^\infty \frac{d \omega}{ \pi} \, \cos (\omega \tau) \,
\tilde \eta(x(t), 0; \omega) , \\ \nonumber
K(t; \tau) &= &\int_0^\infty \frac{d \omega}{2 \pi} \,
\cos(\omega \tau) \, \omega \coth (\omega/2T) \, \tilde \eta (x(t),x(t); \omega),
\end{eqnarray}
with the generalized dissipation 
function $\tilde \eta$ in Eq.~(\ref{BO:tildeEta}).

Before proceeding with the solution of the above stochastic problem,
we briefly address the analytically tractable limit $\Gamma\gg \Delta$.  
While the resulting expressions are not used in the full numerical solution
in Sec.~\ref{sec4}, they are useful to develop intuition
and to determine whether the underdamped vs overdamped
regime is realized, see Sec.~\ref{sss}.  For $\Gamma \gg \Delta$, the Andreev level energy 
is\cite{nazarov,golubov}
\begin{equation}
E_a(x) = \Delta \sqrt{1 - {\cal T}(x) \sin^2 (\phi / 2)}
\end{equation}
with the $x$-dependent effective transmission probability
\begin{equation}
{\cal T}(x) = \frac{1}{1 + \epsilon^2(x)/\Gamma^2} .
\end{equation}
The Andreev level contribution to the fluctuation kernel $K(x(t);\omega)$ 
is then given by
\begin{equation}
K_a(x; \omega) = \frac{\pi\lambda^2}{\Gamma^2}
\sum_{n=0,\pm 1}  \Xi_n(x) \, \delta(\omega - 2 n E_a(x)) .
\end{equation}
With the Fermi function $n_F(\omega)=(e^{\omega/T}+1)^{-1}$, we use the 
auxiliary functions
\begin{eqnarray*}
\Xi_0(x) & = & n_{F}\left(E_a(x)\right) n_F\left(-E_a(x)\right)
 \left( 1 - {\cal T}(x) \right) {\cal T}^3(x) \\ &\times&
\frac{4 \Delta^4}{E_a^2(x)}  \sin^4 (\phi / 2) ,\\
\Xi_{\pm 1}(x) &=&   \left[ n_F\left(\mp E_a(x)\right) \right]^2 {\cal T}^3(x)
\frac{ \Delta^4 \sin^2\phi}{2 E_a^2(x)} .
\end{eqnarray*}
The zero-frequency peak in $K_a(x; \omega)$ is determined by
quantum fluctuations of the Andreev level current\cite{alq} when
the reflectivity is finite, ${\cal T}(x) <1$.
The \textit{continuum} contribution to the above kernels
exhibits only very weak dependence on $\omega$ and
can be evaluated by taking the $\omega \to 0$ limit.  We find
$K_c(x; \omega) \simeq T \eta_c(x; \omega)$ with
$\eta_c(x; \omega) \simeq (2\lambda/\Gamma)^2  e^{-\Delta/T}$.
As a function of $\omega$, both kernels $\eta(x(t);\omega)$ and $K(x(t);\omega)$ 
exhibit a constant background due to the continuum states,
responsible for Ohmic dissipation.\cite{weiss}
Superimposed on this Ohmic part, we have the Andreev
level $\delta$-type contributions. They include a peak at zero
frequency.  We then turn back to the full problem characterized by
arbitrary ratio $\Gamma/\Delta$.

\subsection{Fokker-Planck equation in energy space}

Following standard arguments,\cite{weiss} we can transform
the effective action (\ref{Seff}) to  an equivalent Langevin equation.
Using a Hubbard-Stratonovich transformation,
\[
e^{- y K y/2 } = \int {\cal D} \xi \, e^{-\xi K^{-1} \xi/2 + i \xi y}  ,
\]
we introduce the auxiliary field $\xi(t)$.
Functional integration over $y$ then yields
\begin{equation} \label{calZadiab}
{\cal Z} = \int {\cal D} x \, {\cal D} \xi \,
e^{-\xi K^{-1} \xi/2} \, {\rm det} (\check {\cal G}^{-1})\,
\delta \Big( {\cal L}[x, \xi] \Big) ,
\end{equation}
which enforces the Langevin equation
\begin{equation}
{\cal L}[x, \xi] = \Omega^{-1} \ddot{x} - F(x) + \int^t dt' \,
\eta(t,t')  \dot{x}(t') - \xi(t) =0
\end{equation}
with Gaussian noise $\xi(t)$. The stochastic noise field has zero mean
and variance $\langle \xi(t) \xi(t') \rangle = K(t,t')$.
In what follows, we consider the \textit{weak damping limit},\cite{weiss}
where the damping force ($\propto \dot x$) is small compared to the 
potential force $F(x)$.  In this underdamped regime, the oscillator 
energy $E$ varies slowly on the timescale $\Omega^{-1}$.
We quantify this condition in terms of the system parameters in Sec.~\ref{sss}.

We proceed by writing a Langevin equation for the slow energy variable $E(t)$,
which is averaged in time over the (energy-dependent) oscillator period $T_E$.
Technically, by multiplying the Langevin equation by $\dot x$ and defining
$\tilde \xi = \dot x \, \xi$, we find
\begin{equation} \label{LngvnEq2}
\frac{d}{ dt} \left( \frac{\dot{x}^2 }{2 \Omega} + U(x) \right)  =
- \int^t dt' \, \eta(t,t')  \dot{x}(t) \dot{x}(t') + \tilde \xi(t) ,
\end{equation}
where $U(x)$ is an \textit{effective oscillator potential} with
$F(x) = -\partial_x U(x)$, and
$\langle \tilde \xi(t) \tilde \xi (t')\rangle = K(t,t') \dot x(t) \dot x(t')$.
The change of the oscillator energy 
\[
E(t)= \dot{x}^2/(2\Omega)+U(x)
\]
is thus determined by the work done by the damping force $-\eta \dot{x}$ 
and by the fluctuations.  Averaging Eq.~(\ref{LngvnEq2}) over $T_E$ yields
a Langevin equation in energy space,
\begin{equation} \label{LngvnEq3}
\dot E(t) = - \eta(E) + \xi_E(t) , \quad
\langle \xi_E(t) \xi_E (t') \rangle = K(E)  \delta (t-t') ,
\end{equation}
where $\eta(E)$ determines the energy dissipation rate
and $K(E)$ describes multiplicative (state-dependent) noise,
\begin{eqnarray} \label{kernelsE}
\eta(E) &=& \int_0^{T_E} \frac{dt}{T_E} \int^t dt' \, \eta(t,t')
\, \dot{x}(t) \dot{x}(t'),\\ \nonumber K(E)& =&
 \int_0^{T_E}\frac{dt}{T_E} \int dt' \, K(t,t') \, \dot{x}(t) \dot{x}(t') .
\end{eqnarray}
The corresponding Fokker-Planck equation\cite{weiss}
governing the energy distribution function $w(E,t)$ of the oscillator is
\begin{equation} \label{FPeqE}
\partial_t w(E,t) = \partial_E \left ( \eta(E) w(E,t)  +
\frac12 \partial_E \left[ K(E)  w(E,t) \right] \right) .
\end{equation}
A similar Fokker-Planck equation has been derived
previously\cite{martin,pistolesi} for the normal-state case.
However, the kernels $\eta(t,t')$ and $K(t,t')$
were approximated by time-local [$\propto \delta (t-t')$] 
expressions in Eq.~(\ref{kernelsE}).

The stationary solution of Eq.~(\ref{FPeqE})
is given by the \textit{generalized Boltzmann distribution},
\begin{equation} \label{wE}
w(E) = {\cal N} \,  K^{-1}(E) \, \exp
\left(- \int^E \frac{d E'}{T_{\rm eff}(E')} \right),
\end{equation}
where $T_{\rm eff}(E) = K(E)/2\eta(E)$ is an \textit{effective energy-dependent
temperature} and ${\cal N}$ a normalization constant.

In order to compute the fluctuation-dissipation coefficients (\ref{kernelsE}),
we introduce a velocity-velocity correlation function at energy $E$.
For a periodic solution $x=x(t)=x(t+T_E)$ of
the undamped oscillator problem at given energy $E$, we take the correlator
\begin{eqnarray}\label{QE}
Q_E(t; \tau) &\equiv&  \left . \dot{x}(t + \tau/2) \dot{x}(t - \tau/2) \right|_E \\
& =& \sum_{n = -\infty}^\infty Q_{E,n}(t) \, e^{- i n \Omega_E \tau} ,
\nonumber
\end{eqnarray}
where $Q_{E,n} = Q_{E,n}^\ast = Q_{E,-n}$ and $\Omega_E = 2 \pi / T_E$.
Using Eq.~(\ref{etaX}), we find 
\begin{eqnarray} \label{etaE}
\eta(E) &=& \int_0^{T_E} \frac{dt}{2T_E} \sum_n   Q_{E,n}(t) \,
\tilde \eta(x(t), 0; n \Omega_E),\\ \label{KE}
K(E) &=&  \int_0^{T_E} \frac{dt }{2T_E} \sum_n  Q_{E,n}(t) \\ \nonumber
&\times&  n\Omega_E \coth (n \Omega_E/2T)\, \tilde \eta (x(t),x(t); n\Omega_E).
\end{eqnarray}
While the above equations are straightforward to solve numerically in
the case of a single-well potential, it is also possible to encounter
bistable behavior as reported for the normal-state case.\cite{pistolesi}
We will discuss the transition from a single-well to a double well
potential $U(x)$ in detail in Sec.~\ref{sec4}. 
For the case of a double-well potential $U(x)$ with barrier height $E_b$,
there are two solutions $w_{1,2}(E)$ defined within each well region ($E<E_b$),
and a third solution $w_3(E)$ applicable for energies above
the barrier ($E>E_b$).  These solutions have to be matched by
boundary conditions.\cite{martin,pistolesi} In particular, continuity imposes
$w_1(E_b) + w_2(E_b) = w_3(E_b)$,  and the transition probability to each
well at the separatrix should be equal, $w_1(E_b) = w_2(E_b)$.

\subsection{Current}

In the adiabatic approximation, the Josephson current is given by\cite{zazu1}
\begin{equation}
I  = - \Delta \sin (\phi/2)  \int\frac{d \omega}{2 \pi i} \,
f(\omega)  \, \alpha_\omega \, {\rm Tr}_N \, \left(
\sigma_x {\langle {\cal G}^R(x; \omega)\rangle}_{\rm osc} \right) ,
\end{equation}
which involves time-averaging over an oscillator period $T_E$ for given $E$,
followed by an average over the oscillator energy using the 
stationary distribution (\ref{wE}), 
\begin{eqnarray} \label{avrgosc}
{\langle \check {\cal G}(x; \omega) \rangle}_{\rm osc} &=& \int dE \,  w(E) \,
\int_0^{T_E} \frac{dt}{T_E} \\ \nonumber &\times&
\delta \left( \frac{\dot{x}^2}{2 \Omega} + U(x) - E \right) \,
\check {\cal G}(x(t); \omega) .
\end{eqnarray}
Analytic continuation then yields for the Josephson current
\begin{equation}\label{joscur}
I(\phi)=- 2 T \Delta^2 \sin (\phi) \sum_{\nu_n > 0}
\tilde \alpha_{\nu_n}^2 \, \left\langle {\cal D}^{-1}(x; i \nu_n)
\right\rangle_{\rm osc}
\end{equation}
with fermion Matsubara frequencies $\nu_n = (2 n + 1) \pi T$ (integer $n$).
Equation (\ref{alphadef}) yields
$\tilde \alpha_\nu = \alpha_{i\nu} = \Gamma/\sqrt{\Delta^2 + \nu^2}$,
and a similar result is obtained for ${\cal D}$ from Eq.~(\ref{calD}).

\subsection{Underdamped regime}\label{sss}

In practice, the physically most relevant parameter regime 
corresponds to underdamped motion of the oscillator.  This can be shown by an
estimate for $\eta(E)$ given next.  In Sec.~\ref{sec4}, we also show the 
full numerical result for $\eta(E)$ to self-consistently
verify that one indeed stays in the weak-damping limit.
Our analytical estimates were obtained for $\Gamma/\Delta\gg 1$.

For given energy $E$, the Andreev level contribution $\eta_a(E)$ 
is non-zero only if the oscillator path $x = x_E(t)$ 
passes through $x = 0$. We find 
\begin{equation}\label{etaEa}
\frac{\eta_a(E)}{ \Omega^2} \simeq g_a \sqrt{E / \Omega} 
\end{equation}
with the dimensionless number
\[
g_a = n_F(E_0) n_F(-E_0) \frac{\lambda \epsilon_0}{T E_0}
\left( \frac{{\cal T}(0) \Delta \sin(\phi/2)}{ \Gamma }\right)^2  ,
\]
where $E_0 = E_a(0)$ denotes the bare Andreev level energy.
This estimate is obtained for zero Andreev level width $\gamma_a = 0$
and by neglecting the ($E$-dependent) renormalization of the
 oscillator frequency.  (In the numerical analysis below, we use
$\gamma_a = 0.01 \Delta$.)

On the other hand, the continuum contribution to the damping kernel $\eta$ 
is estimated by
\begin{equation}\label{etaEc}
\frac{\eta_c(E)}{ \Omega^2 } \simeq \frac{T E}{\Omega \Delta}
 \left( \frac{2 \lambda}{\Gamma} \right)^2  e^{-\Delta/T}.
\end{equation}
Note that the two contributions scale differently with $E$. 
The underdamped regime is realized when $\eta(E)/\Omega^2<1$.  
It is straightforward to observe from the above expressions 
that for $\lambda\alt \Gamma$, this condition is always
fulfilled. The underdamped regime may cover even
significantly larger electron-vibration couplings $\lambda$.  

\section{Results and discussion}
\label{sec4}

Let us now describe results obtained from this semiclassical approach.  
We here only consider parameter sets consistent with
the assumption of \textit{underdamped adiabatic motion} of the oscillator,
see Sec.~\ref{sss}.  In addition, we shall assume good coupling between dot and electrodes,
$\Gamma / \Delta >  1$, consistent with the fact that we 
neglect Coulomb interaction effects on the dot.  Note that 
the opposite case $\Gamma / \Delta < 1$ was studied in Ref.~\onlinecite{zazu2sq}.

The numerical calculation goes as follows.
We first compute the effective potential $U(x)$ according to Eq.~(\ref{Fbath2}).
Having determined the effective potential $U(x)$, the calculation proceeds
by computing $Q_{E,n}$ as defined in Eq.~(\ref{QE}).  This involves a numerical
solution of the classical equations of motion in the potential $U(x)$,
which are always periodic (the oscillation period $T_E$ is thereby obtained 
numerically).  Subsequently we compute the damping kernel $\eta(E)$ using Eq.~(\ref{etaE}),
and the fluctuation kernel $K(E)$ from Eq.~(\ref{KE}).
These kernels then result in the probability distribution $w(E)$ according
to Eq.~(\ref{wE}), and finally the Josephson current-phase
relation is obtained from Eq.~(\ref{joscur}).

\begin{figure}
\scalebox{0.35}{\includegraphics{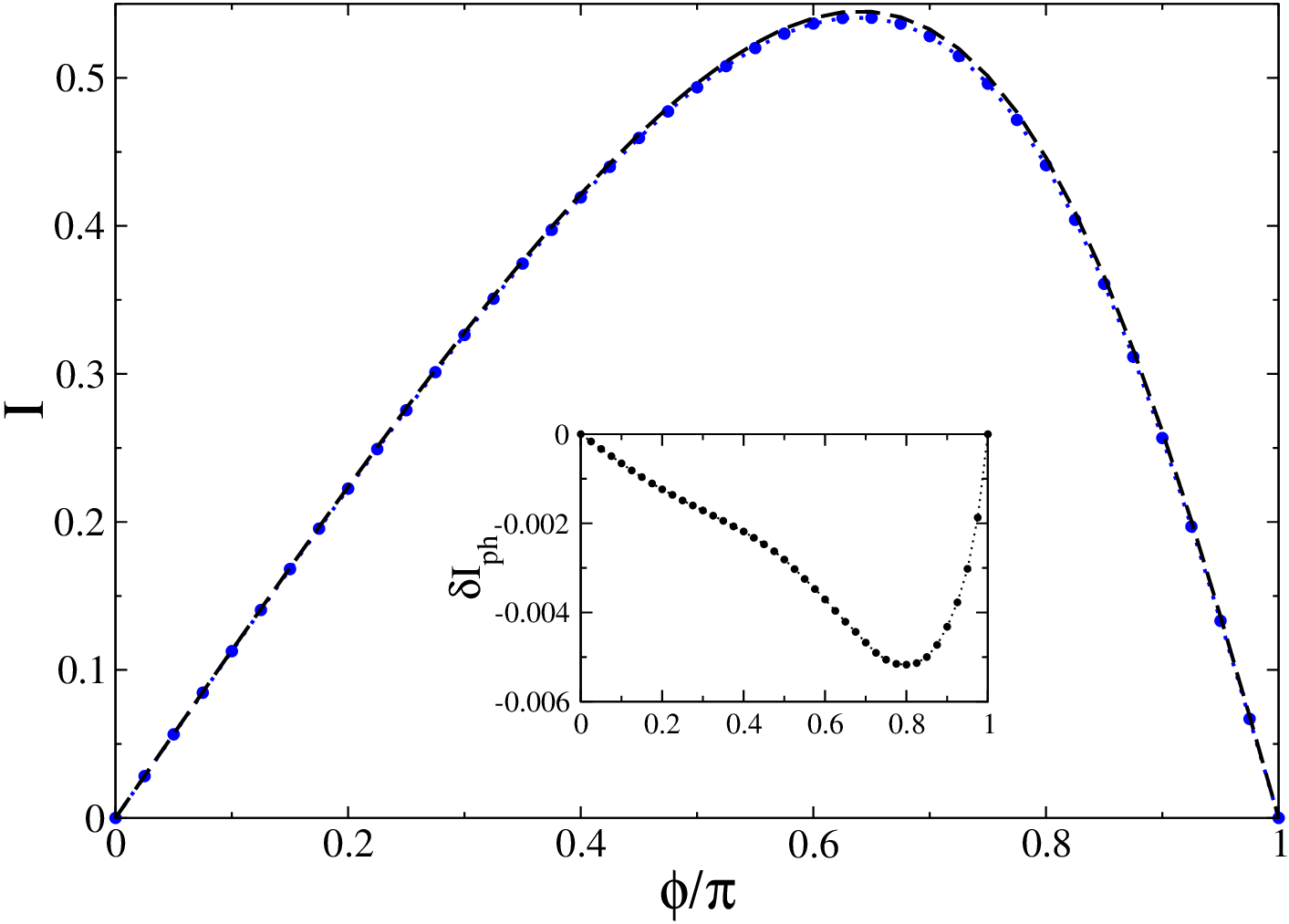}}
\caption{\label{fig1} (Color online)
Josephson current in units of $e\Delta/\hbar$ vs superconducting phase
difference $\phi$ for the interacting case ($\lambda=0.5\Delta$: blue circles) and 
for $\lambda=0$ (dashed curve). The system parameters (in units of $\Delta$) are
 $\Gamma=8, \epsilon_0=-0.1, \Omega=0.05,$ with temperature $T=0.2$.
 Inset: Interaction correction to the current, $\delta I_{\rm ph}= I(\lambda)-I(\lambda=0)$,
vs phase difference $\phi$ for the data in the main panel.  }
\end{figure}
\begin{figure}
\scalebox{0.35}{\includegraphics{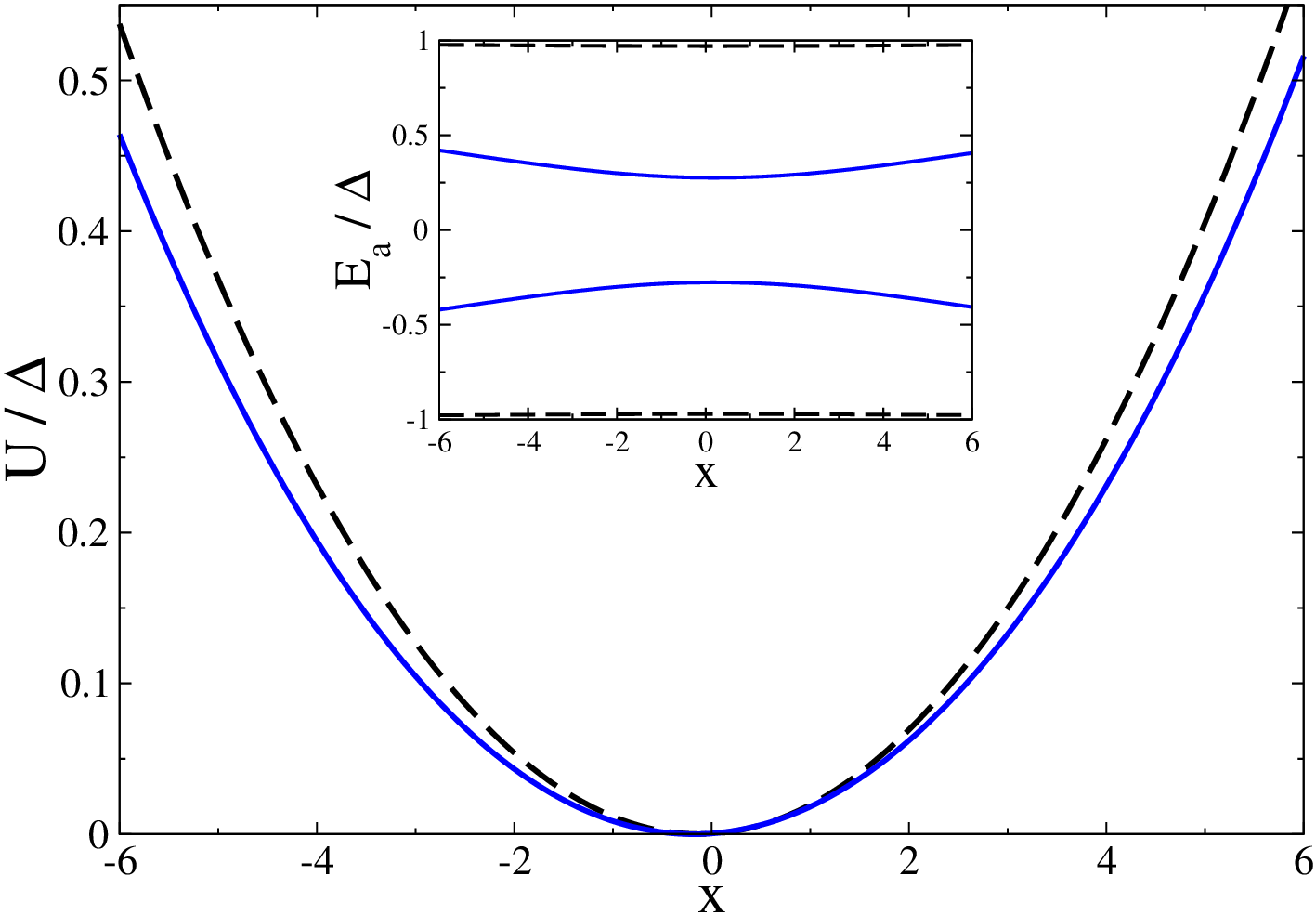}}
\caption{\label{fig2} (Color online)
Effective potential $U(x)$ vs dimensionless oscillator coordinate $x$ for
$\phi = 0$ (black dashed) and $\phi=0.8 \pi$ (blue solid curve).
System parameters are as in Fig.~\ref{fig1}.  Inset: Andreev level spectrum 
vs $x$ for the two quoted values of $\phi$.  }
\end{figure}
\begin{figure}
\scalebox{0.35}{\includegraphics{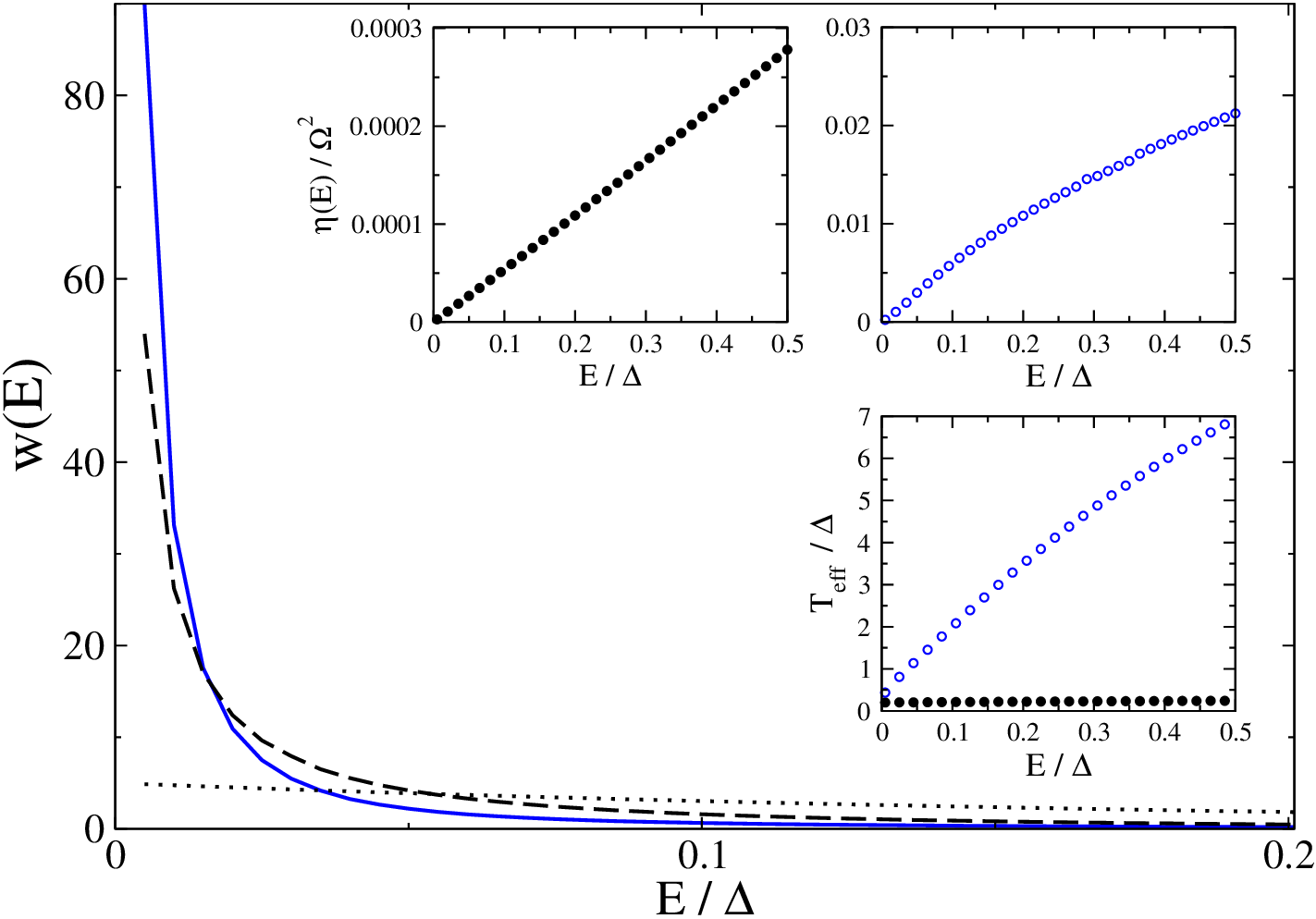}}
\caption{\label{fig3} (Color online)
Energy distribution function (\ref{wE}) for $\phi=0$ (black dashed)
and $\phi = 0.8 \pi$ (blue solid curve), using the same parameter
set as in Fig.~\ref{fig1}.  The dotted curve shows a Boltzmann
distribution for temperature $T$.  Insets: 
Damping kernel $\eta(E)$ vs energy $E$ for $\phi=0$ [top left] and for
$\phi=0.8 \pi$ [top right]. The corresponding effective 
temperature $T_{\rm eff}=K(E)/2\eta(E)$ is shown as a function of $E$
for these two values of $\phi$ in the bottom right inset:
black filled circles are for $\phi=0$, and blue open
circles are for $\phi=0.8\pi$.}
\end{figure}

\subsection{Single-well case}

Figures \ref{fig1}, \ref{fig2} and \ref{fig3}
show our numerical results for the following set of system parameters:
$\Gamma = 8 \Delta$, $\epsilon_0 = -0.1 \Delta$, $\Omega = 0.05 \Delta$,
$\lambda=0.5 \Delta$, with temperature put to $T=0.2 \Delta$.
For this parameter set, we are in the weak-coupling (underdamped) regime,
where the above formalism can be safely applied.
The effective potential $U(x)$ then has a single minimum
for all values of the phase difference $\phi$, see Fig.~\ref{fig2} for
$\phi=0$ and $\phi=0.8\pi$.  This single-well behavior of the 
effective oscillator potential surface can be rationalized by noting that 
the electron force $F_e(x)$ is here mainly 
determined by the continuum contribution,
which in turn is almost insensitive to the phase difference $\phi$.
Interestingly, as seen in Fig.~\ref{fig1}, 
the Josephson current is basically not modified,
with only a very small negative interaction 
correction even for a relatively strong electron-vibration coupling $\lambda$.
The weak sensitivity of the current to $\lambda$ comes from a
strong localization of the oscillator near the 
bottom of the effective potential $U(x)$ at $x = 0$, see Fig.~\ref{fig2}.  

The corresponding distribution functions $w(E)$ for 
$\phi = 0$ and $\phi=0.8 \pi$ are shown in Fig.~\ref{fig3}.
The observed singular behavior for small energies $E$ 
is mainly determined by the factor $K^{-1}(E)$ in Eq.~(\ref{wE}).
For instance, for the continuum contribution, one obtains
$K_c(E) \simeq 2 T \eta_c(E) \propto E$, and hence we find the 
scaling $w(E) \propto 1/E$.
The approximately linear law $K(E)\propto E$ as $E\to 0$ stays also valid
when including the Andreev level contribution.
Indeed, from Eq.~(\ref{KE}), we  find $K(E) \approx E T  \overline{\tilde \eta (x,x)}$,
with the average over phase space (at given energy $E$) defined as
$\overline{\tilde \eta (x,x)} = \frac{\oint dx \, p(x) \; \tilde \eta (x,x)
}{  \oint dx \, p(x) }.$  
Note that $\oint dx \, p(x) \approx 2 \pi E / \Omega$, while
$\overline{\tilde \eta (x,x)}$ is only weakly dependent on $E$. 
We thus conclude again that $K(E) \propto E$.

For small $\phi$, the main contribution to the oscillator damping
comes from the continuum states, while for intermediate $\phi$
the Andreev level contribution starts to dominate.
This is explicitly seen in the two upper insets in Fig.~\ref{fig3}, where $\eta(E)$
is shown for $\phi = 0$ and $\phi=0.8 \pi$, respectively.
For $\phi=0$, we find a linear $E$-dependence, which turns
into a square-root dependence for $\phi=0.8\pi$,
in accordance with Eqs.~(\ref{etaEc}) and (\ref{etaEa}), respectively.
Furthermore, the bottom right inset of Fig.~\ref{fig3} shows that
the effective temperature $T_{\rm eff}(E) = K(E)/2\eta(E)$ 
is greatly enhanced for $\phi=0.8\pi$ due to Andreev-level current 
fluctuations. These fluctuations  lead to stronger localization of the 
oscillator at low $E$.
For $\phi = 0$, we find $T_{\rm eff}(E)\simeq T$, as expected
when the continuum contributions dominate.  Nevertheless, even then the 
oscillator distribution function $w(E)$
strongly deviates from the classical Boltzmann distribution of a free oscillator
(shown in Fig.~\ref{fig3} for comparison). This difference can be traced to
the prefactor $K^{-1}(E)$ in Eq.~(\ref{wE}).

\begin{figure}
\scalebox{0.35}{\includegraphics{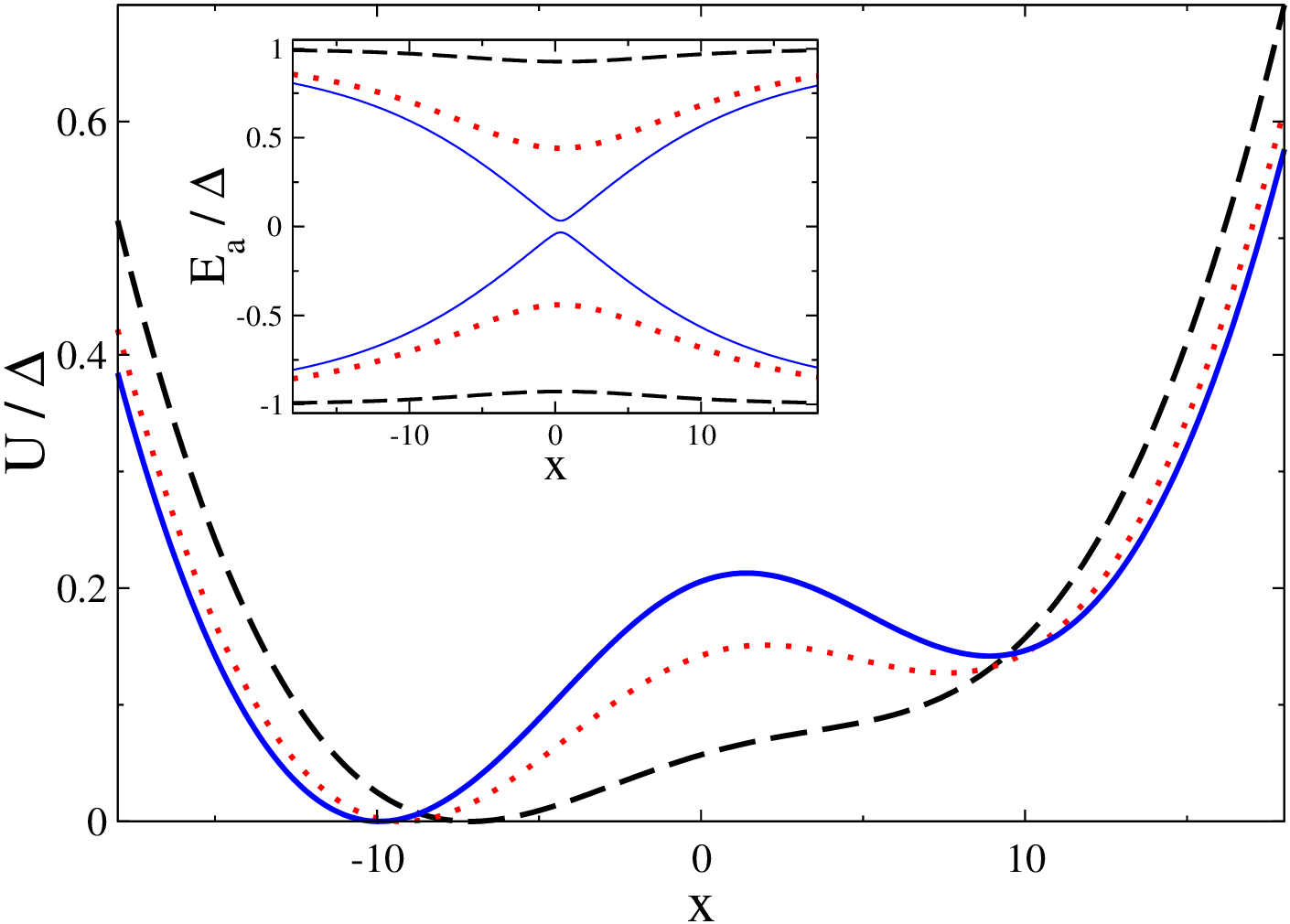}}
\caption{\label{fig4} (Color online)
Effective potential $U(x)$ vs $x$ for
$\phi = 0$ (black dashed), $\phi=0.65 \pi$ (red dotted), and $\phi= 0.975\pi$
(blue solid curve). System parameters (with $\Delta=1$)
are $\Gamma = 4.8, \epsilon_0 = -0.15, \Omega = 0.02, \lambda=0.4, T=0.25$.
Inset: Andreev level spectrum vs $x$ for these three values of $\phi$.}
\end{figure}
\begin{figure}
\scalebox{0.35}{\includegraphics{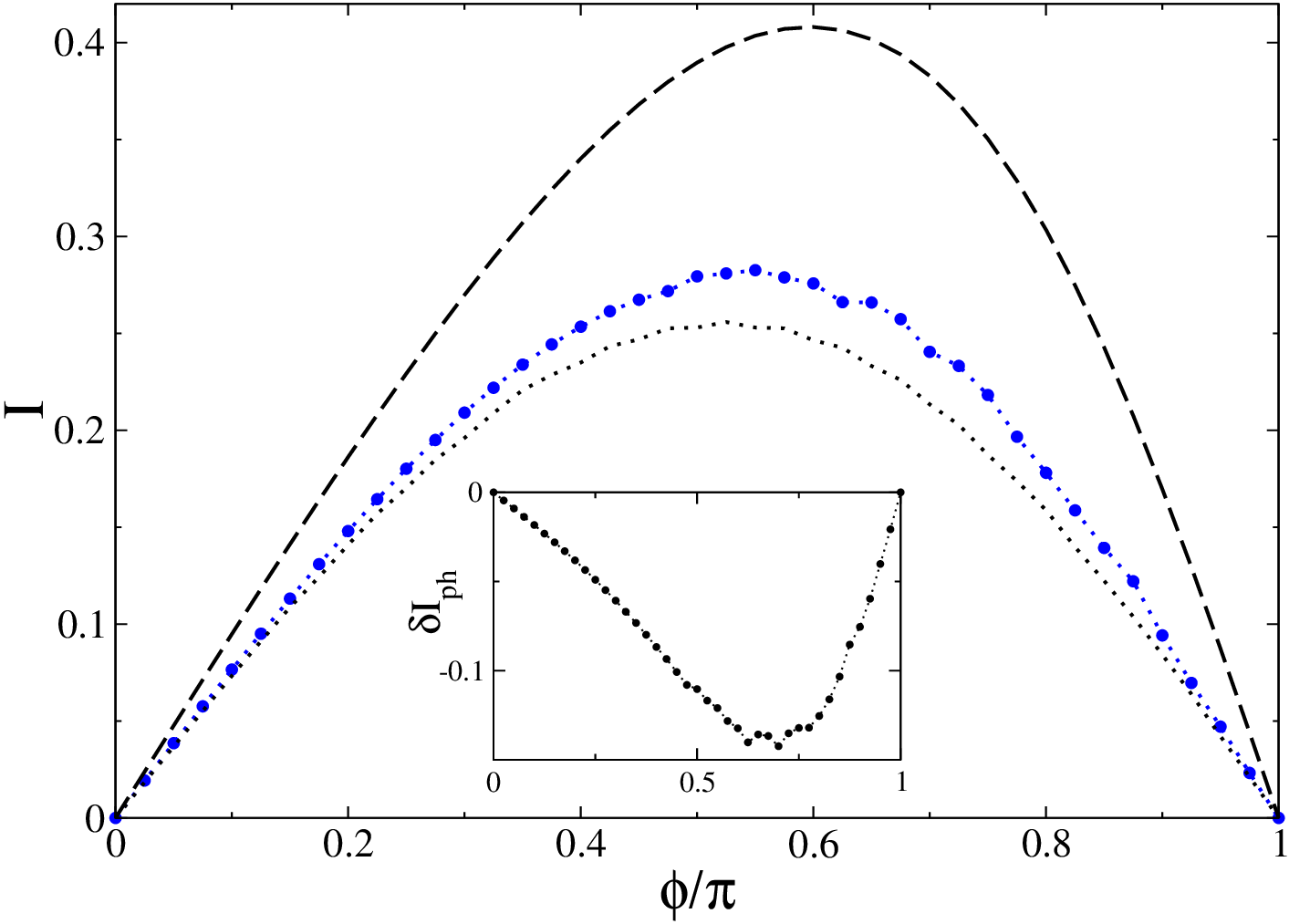}}
\caption{ \label{fig5}
(Color online) Josephson current (in units of $e\Delta/\hbar$) vs
$\phi$ for the interacting case ($\lambda=0.4\Delta:$ blue circles)
and for $\lambda=0$ (black dashed curve). The thin dotted curve shows the result
when $x$ is held fixed at the global minimum of $U(x)$. System parameters are
as in Fig.~\ref{fig4}.
Inset: Interaction correction $\delta I_{ph}$ vs phase difference.}
\end{figure}

\subsection{Crossover to the double-well potential}

Let us next analyze a second parameter set, where we will encounter a 
nontrivial double-well behavior for the effective
 oscillator potential $U(x)$.  The
transition from single- to double-well behavior is here induced by a variation
of the phase difference $\phi$, and one may therefore affect
the conformational state of the molecule in a dissipationless manner 
in such a setup.  The parameter set is given by 
$\Omega=0.02\Delta$, $\Gamma=4.8\Delta$,
$\epsilon_0=-0.15\Delta$, with electron-vibration coupling strength
$\lambda=0.4\Delta$.  Moreover, the temperature has been set to $T=0.25\Delta$.
As illustrated in Fig.~\ref{fig4}, we indeed find a
transition between a single- and a double-well potential
induced by a variation of $\phi$. 
Similar transitions (with associated bistabilities) were
reported for a two-level system instead of the oscillator,\cite{zazu3} 
and for the nonequilibrium normal-state local Holstein model.\cite{pistolesi}

Although the continuum contribution to the electron
force and thus to the effective potential $U(x)$
still plays an overall dominant role, it is almost insensitive to
variations of $\phi$.  The $\phi$-tunable transition to a double-well
potential shown in Fig.~\ref{fig4} is therefore caused by Andreev level
contributions to $F_e(x)$.
We note that the shape of $U(x)$ is also sensitive to temperature
through thermal occupation factors of the Andreev levels.
The dynamical frequency $\Omega_E$ for the oscillator motion in the 
effective potential $U(x)$ can  be strongly renormalized away from 
the bare oscillator frequency $\Omega$.
For the parameters in Fig.~\ref{fig4}, we typically
find $\Omega_E \approx 0.5\Omega$.  The $x$-dependence of the
Andreev level spectrum $E_a(x)$ in the adiabatic
limit, i.e., with instantaneous $x=x(t)$, is shown for several
phases $\phi$ in the inset of Fig.~\ref{fig4}.
Note that this spectrum is rather different from the featureless
Andreev level spectrum for the first parameter set, see 
inset of Fig.~\ref{fig2}.

Figure \ref{fig5} shows the current-phase relation for this parameter set.
The Josephson current again exhibits an overall suppression due to the
electron-vibration coupling as reported
previously.\cite{novotny,zazu1,bcs-holstein1,schattka} 
The suppression is now more pronounced than in Fig.~\ref{fig1}, 
but still remains moderate.
Moreover, the current-phase relation exhibits small yet 
characteristic cusps in the
crossover region between the
 single- and double-well situation ($\phi\approx 0.6\pi$ to $0.7\pi$),
where Andreev level noise $\propto I^2(\phi)$ is strongly enhanced.\cite{alq,LY}
However, the effect of switching between the two potential wells
does not have a dramatic influence on the current-phase relation
because the magnitude of the current is basically the same in each well:
the coordinates $x_{1,2}$ of two local minima are almost symmetric
with respect to $x=0$. Indeed, we find $x_1 \approx -x_2$
and hence $\epsilon(x_1) \approx \lambda x_1 \approx -\epsilon(x_2)$
for strong coupling $\lambda$ and small $\epsilon_0$.
\begin{figure}
\scalebox{0.35}{\includegraphics{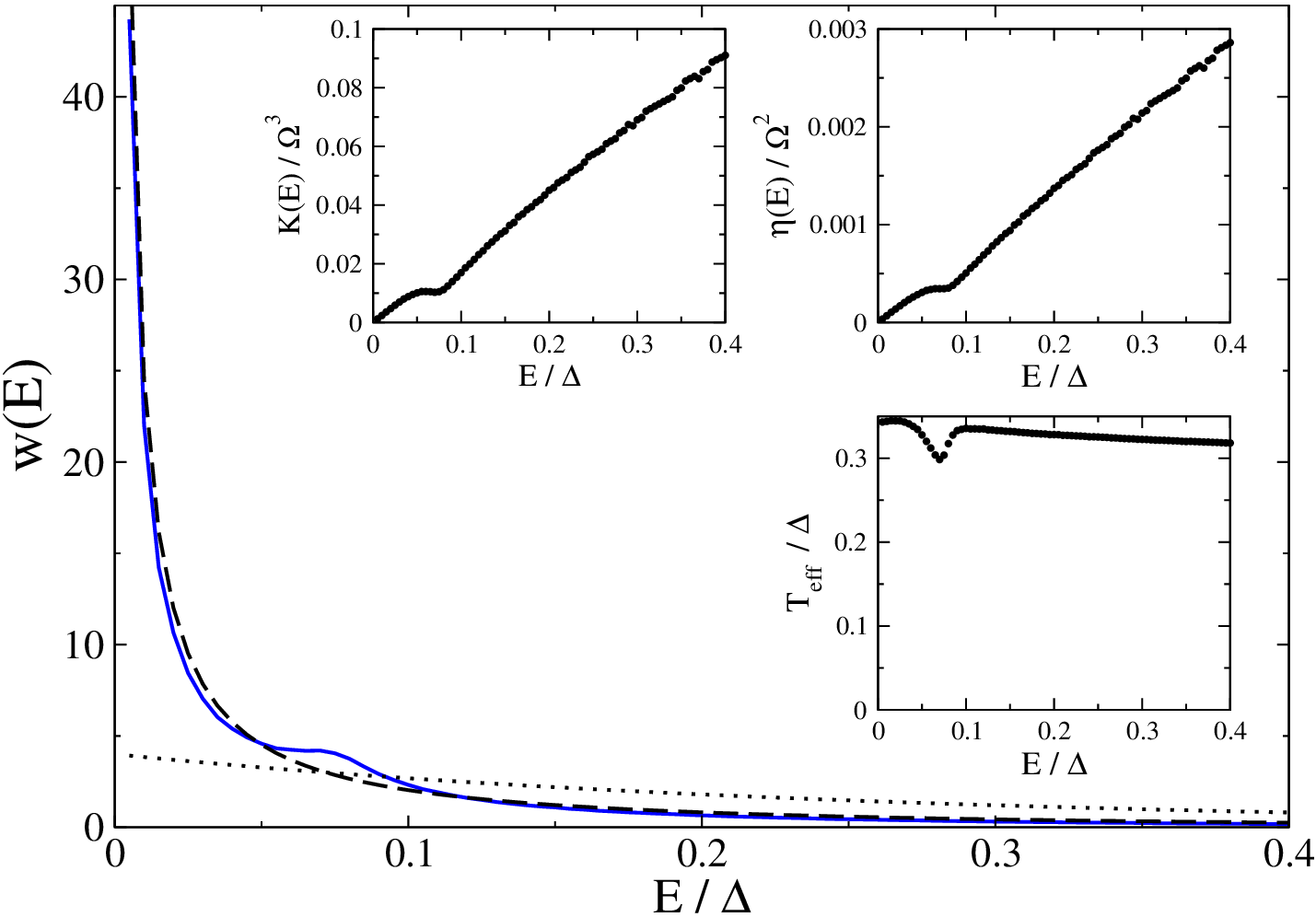}}
\caption{ \label{fig6}
(Color online) Energy distribution function (\ref{wE}) (blue solid curve) for $\phi=0$
with parameters in Fig.~\ref{fig4}.  Note that the potential $U(x)$ then
has a single minimum.  For comparison, the black dotted (black dashed)
curve shows a Boltzmann (Bose-Einstein) distribution function for the given
temperature $T=0.25\Delta$.  The three insets show the corresponding
energy-dependence of the kernels $K(E)$ [top left],
the damping kernel $\eta(E)$ [top right],  and
the effective temperature $T_{\rm eff}(E)=K(E)/2\eta(E)$ [bottom right].  }
\end{figure}
\begin{figure}
\scalebox{0.35}{\includegraphics{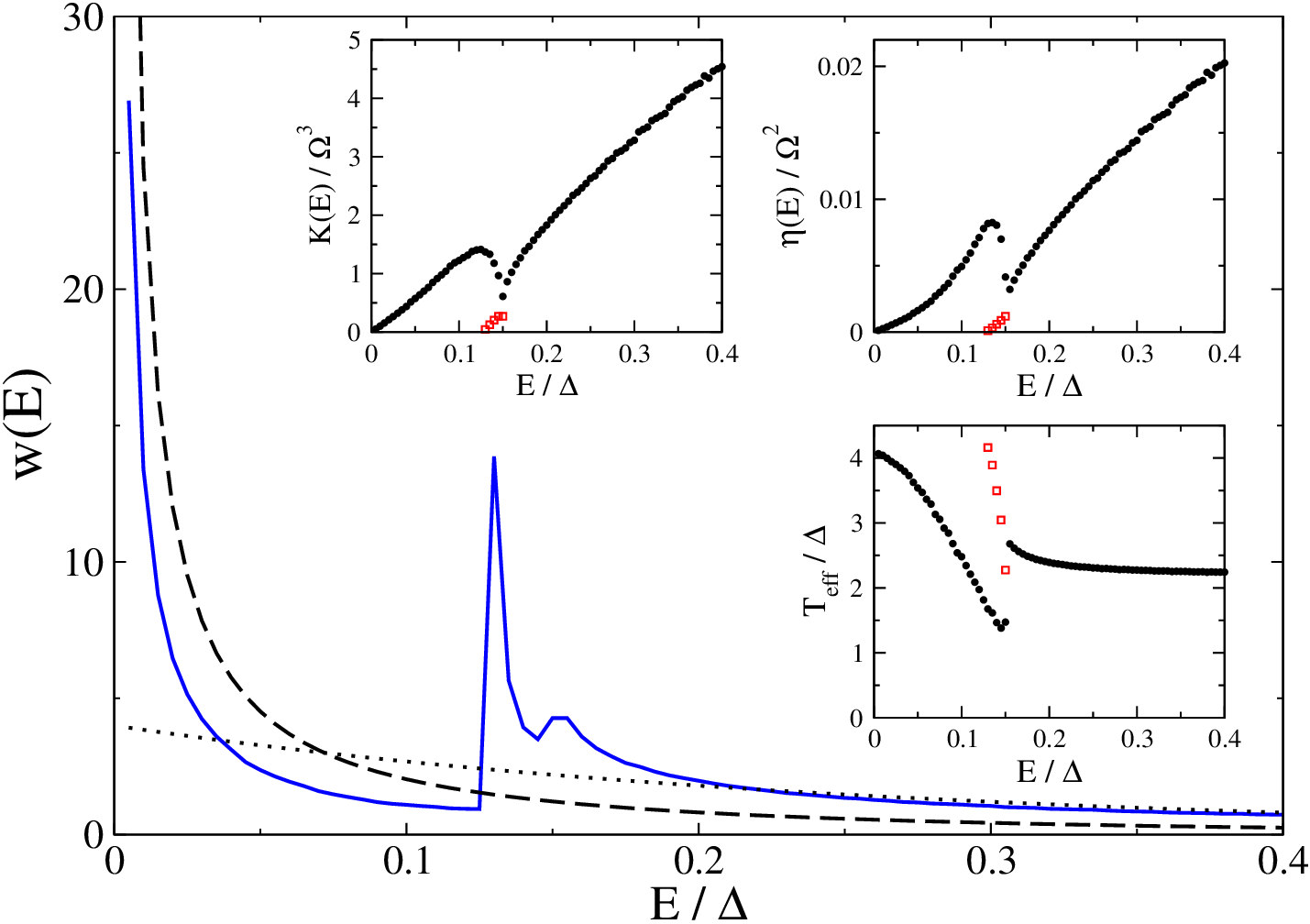}}
\caption{ \label{fig7} (Color online)
Same as Fig.~\ref{fig6} but for $\phi= 0.65\pi$ (crossover regime
between single- and double-well potential).
In the insets, results for the deeper and the shallower well below the potential barrier
are shown by filled black circles and open red squares, respectively.  }
\end{figure}
\begin{figure}
\scalebox{0.35}{\includegraphics{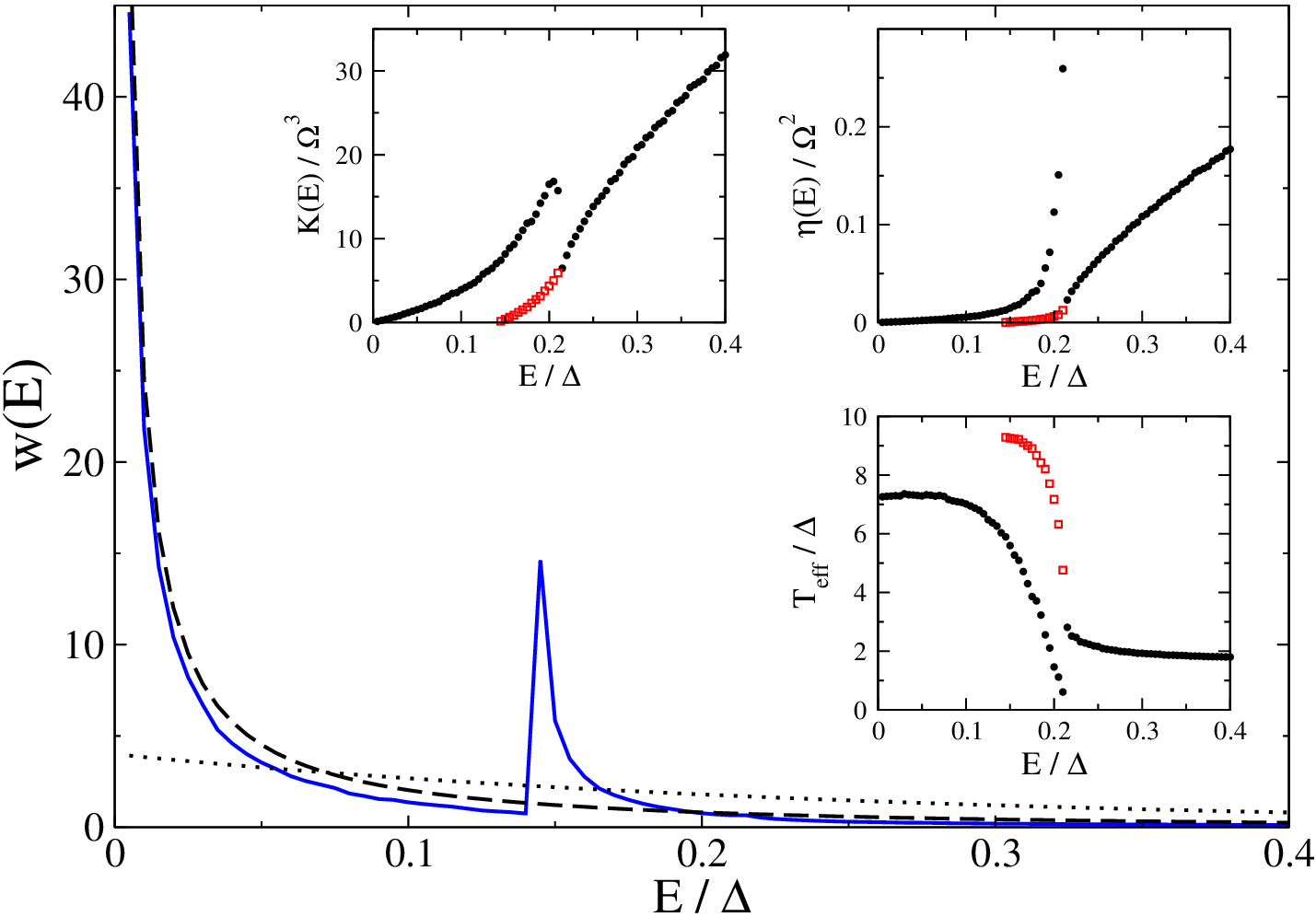}}
\caption{ \label{fig8} (Color online)
Same as Fig.~\ref{fig7} but for $\phi = 0.975\pi$, where
$U(x)$ in Fig.~\ref{fig4} corresponds to a double-well potential.}
\end{figure}

Figures \ref{fig6}, \ref{fig7} and \ref{fig8}
then show the resulting energy distribution function $w(E)$
for the three values of the phase difference $\phi$ considered in Fig.~\ref{fig4},
 respectively.  Despite of the \textit{enhanced} effective temperature
$T_{\rm eff}(E)$ as compared to $T$, we find again that $w(E)$ strongly
deviates from a classical (Boltzmann) distribution function.
Interestingly, for energies $E$ below the separatrix region,
$w(E)$ is well approximated by an effective Bose-Einstein function.
Indeed, we find that for $E \ll T$, both $w(E)$ and the Bose-Einstein function $n_B(E)$
obey the same equation.  As a result, within this region, we find $w(E)
\propto 1/E$ instead of the Boltzmann dependence $\propto e^{-E/T}$,
implying a much stronger localization of the oscillator's motion
near the bottom of the deeper well.  Moreover, the damping $\eta(E)$ is
determined by both the continuum and Andreev level contributions,
while the diffusion coefficient $K(E)$ is essentially determined
by the Andreev level contribution.

\begin{figure}
\scalebox{0.35}{\includegraphics{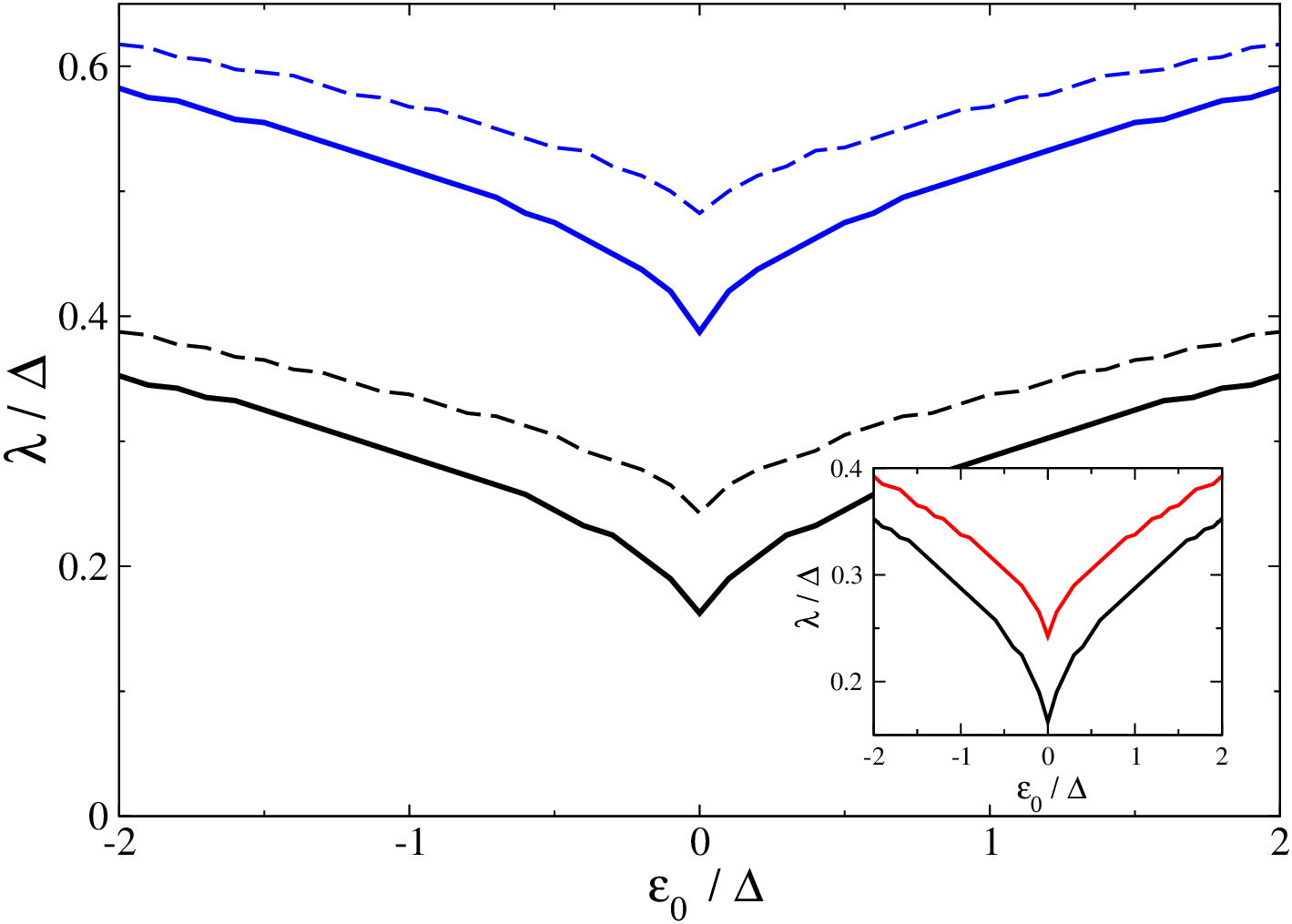}}
\caption{ \label{fig9} (Color online)
Effective phase diagram in the $\lambda-\epsilon_0$ plane for 
$\Omega=0.02$ (units are such that $\Delta=1$). The main panel
shows $\lambda_{c1}(\epsilon_0)$ for $T=0.1$ (solid curves) and
$T=0.4$ (dashed curves),
where single-well behavior is found for $\lambda<\lambda_{c1}$
and double-well behavior starts to set in for $\lambda>\lambda_{c1}$.
The black (lower) curves are for $\Gamma = 2$, and the blue (upper)
curves are for $\Gamma = 8$.
Inset: Boundaries $\lambda_{c1}$ (black lower curve) and $\lambda_{c2}$ (red
upper curve) 
for $\Gamma=2$ and $T=0.1$. For $\lambda>\lambda_{c2}$, the 
double-well behavior occurs for all values of the phase difference $\phi$.
}
\end{figure}

Finally, we address the parameter regime where the described switching from 
single- to double-well behavior in $U(x)$ is found, see Fig.~\ref{fig9}.
For simplicity, we consider a fixed vibrational frequency, $\Omega=0.02\Delta$.
For given system parameters, when increasing $\lambda$, we find from
our numerical scheme that
double-well behavior starts to appear at $\lambda = \lambda_{c1}$ for
$\phi = \pi$.  When further increasing $\lambda$, the double-well behavior
extends to a region with $\phi<\pi$ as well. 
A second scale $\lambda_{c2}>\lambda_{c1}$ is then 
defined such that for $\lambda \ge \lambda_{c2}$,
 the double-well behavior is found for all $\phi$.
In order to determine $\lambda_{c1,2}$, it is therefore sufficient
to probe for the single-to-double-well transition at the phase differences
$\phi = \pi$ and $\phi= 0$. The transition region 
between $\lambda_{c1}$ and $\lambda_{c2}$ is in fact rather narrow,
cf.~the inset of Fig.~\ref{fig9}.
Note that the location of the switching transition 
is quite sensitive to temperature. In particular,
with increasing $T$, the boundary $\lambda_{c1,2}$
shifts to bigger $\lambda$ values.

\section{Conclusions} \label{conc}

We have shown that the adiabatic limit allows to make analytical progress 
for an important model of molecular electronics, 
the superconducting local Holstein model.
It describes a spinless resonant electronic level 
coupled to a single boson (vibration
mode), where the resonant level is coupled to two 
superconducting reservoirs with
a phase difference $\phi$.  The adiabatic limit is realized when the oscillator
frequency $\Omega$ is smaller than both the superconducting gap $\Delta$ and the
dot-to-lead hybridization energy scale $\Gamma$.  This regime allows for a 
semiclassical Born-Oppenheimer-type treatment, where the electronic degrees of 
freedom can be integrated out and give rise to an effective oscillator potential
$U(x)$.  Moreover, they cause dissipative damping and a stochastic noise force.
The most relevant parameter regime turns out to be the underdamped one, where it
is appropriate to consider diffusion in energy space, and the effects of 
damping $[\eta(E)]$ and noise $[K(E)]$ can be taken into account within a
standard Fokker-Planck scheme. The resulting distribution function $w(E)$
solving the Fokker-Planck equation can be obtained numerically with moderate effort,
and allows us to obtain quantitative results 
within a controlled approximation for $\Omega\ll {\rm min}(\Delta,\Gamma)$.

The method has been applied to a calculation of the Josephson current-phase
relation $I(\phi)$.  While the resulting corrections to the Josephson current
are generally small in magnitude even for strong 
electron-vibration coupling $\lambda$,
they still cause some features when the effective potential $U(x)$ 
changes character.  In particular, it is possible to induce a change from a
 single- to a double-well potential surface by variation of the 
phase difference $\phi$. Near the transition point, we predict enhanced Andreev 
level noise.  Similar transitions from single- to double-well effective 
potentials were also reported for the normal-state case,\cite{pistolesi} 
but for a nonequilibrium situation
where a finite bias voltage is applied and dissipation is unavoidable. 
In our equilibrium case, the transition is induced by a variation of 
the superconducting phase and therefore is dissipationless. 
However, the effect of switching from one to two minima in $U(x)$ 
on the Josephson current $I(\phi)$ is much weaker here, which can be
rationalized by noting that the two minima are located symmetrically
and the Josephson current is essentially identical when the oscillator
is close to a given minimum.  It is also worth mentioning that the 
Josephson current seems always to be \textit{suppressed} by the 
coupling to the oscillator,
independent of the normal-state transmission probability through the junction,
i.e., the interaction correction is negative, in contrast to what
happens in the normal state.\cite{egger}   This conclusion has also
been reached via perturbation theory in the
 electron-vibration coupling $\lambda$
for the Josephson current.\cite{zazu1,schattka}

A quantity that is much more sensitive to the existence of two minima
in the effective oscillator potential $U(x)$ is the phonon distribution
function.  As we have discussed in Sec.~\ref{sec4}, in the double-well
case, the phonon distribution function has a characteristic two-peak 
structure and displays strong phonon localization. 
It may be possible to access this quantity experimentally via
resonant coherent phonon spectroscopy techniques,\cite{gambetta,phononexp} 
and thereby provide
clear signatures of the predicted crossover from single- to double-well
behavior.

\acknowledgments
This work was supported by the SFB TR 12 of the DFG and by the EU
network INSTANS.


\begin{thebibliography}{10}

\bibitem{nazarov}
Yu.V. Nazarov and Ya.M. Blanter, \textit{Quantum transport} (Cambridge University Press,
Cambridge, 2009).

\bibitem{nitzan}
A. Nitzan and M.A. Ratner, Science \textbf{300}, 1384 (2003).

\bibitem{boese}
D. Boese and H. Schoeller, Europhys. Lett. {\bf 54}, 668 (2001).

\bibitem{flensberg}
K. Flensberg, Phys. Rev. B {\bf 68}, 205323 (2003);
S. Braig and K. Flensberg, {\sl ibid.} {\bf 68}, 205324 (2003).

\bibitem{cornaglia}
P.S. Cornaglia, H. Ness, and D.R. Grempel, Phys. Rev. Lett. {\bf 93},
147201 (2004).

\bibitem{mitra}
A. Mitra, I. Aleiner, and A.J. Millis, Phys. Rev. B {\bf 69}, 245302 (2004).

\bibitem{paulsson}
M. Paulsson, T. Frederiksen, and M. Brandbyge,
Phys. Rev. B {\bf 72}, 201101(R) (2005).

\bibitem{koch}
J. Koch and F. von Oppen, Phys. Rev. Lett. {\bf 94}, 206804 (2005).

\bibitem{martin}
D. Mozyrsky, M.B. Hastings, and I. Martin,
Phys. Rev. B  {\bf 73}, 035104 (2006).

\bibitem{zazu2sq}
A. Zazunov, D. Feinberg, and T. Martin, Phys.  Rev. Lett. \textbf{97}, 196801 (2006).

\bibitem{zazu2}
A. Zazunov, D. Feinberg, and T. Martin, Phys. Rev. B {\bf 73}, 115405 (2006);
A. Zazunov and T. Martin, \textit{ibid.} {\bf 76}, 033417 (2007).

\bibitem{richter}
A. Donarini, M. Grifoni, and K. Richter, Phys. Rev. Lett.
{\bf 97}, 166801 (2006).

\bibitem{egger}
R. Egger and A.O. Gogolin, Phys. Rev. B {\bf 77}, 113405 (2008).

\bibitem{lothar}
L. M\"uhlbacher and E. Rabani, Phys. Rev. Lett. {\bf 100}, 176403 (2008).

\bibitem{marten}
M. Leijnse and M.R. Wegewijs, Phys. Rev. B {\bf 78}, 235424 (2008).

\bibitem{pistolesi}
F. Pistolesi, Ya.M. Blanter, and I. Martin, Phys. Rev. B {\bf 78},
085127 (2008).

\bibitem{fabrizio} P. Lucignano, G.E. Santoro, M. Fabrizio, and E.
Tosatti, Phys. Rev. B \textbf{78}, 155418 (2008).

\bibitem{nitzan2}
M. Galperin, M.A. Ratner, and A. Nitzan,
J. Phys. Cond. Matt. {\bf 19}, 103201 (2007).

\bibitem{organic}
N.B. Zhitenev, H. Meng, and Z. Bao, Phys. Rev. Lett. {\bf 88}, 226801 (2002);
X.H. Qiu, G.V. Nazin, and W. Ho, {\sl ibid.} {\bf 92}, 206102 (2004);
L.H. Yu, Z.K. Keane, J.W. Ciszek, L. Cheng, M.P. Stewart, J.M. Tour, and D.
Natelson, {\sl ibid.} {\bf 93}, 266802 (2004).

\bibitem{park}
H. Park, J. Park, A.K.L. Lim, E.H. Anderson, A.P. Alivisatos,
and P.L. McEuen, Nature {\bf 407}, 57 (2000).

\bibitem{natelson}
J. Park, A.N. Pasupathy, J.I. Goldsmith, C. Chang, Y. Yaish, J.R. Petta,
M. Rinkoski, J.P. Sethna, H.D. Abruna, P.L. McEuen, and D.C. Ralph,
Nature {\bf 417}, 722 (2002);
L.H. You and D. Natelson, Nano Lett. {\bf 4}, 79 (2004).

\bibitem{paul}
A.N. Pasupathy, J. Park, C. Chang, A.V. Soldatov, S. Lebedkin,
R.C. Bialczak, J.E. Grose, L.A.K. Donev, J.P. Sethna,
D.C. Ralph, and P.L. McEuen,  Nano Lett. {\bf 5}, 203 (2005).

\bibitem{leroy}
B.J. LeRoy, S.G. Lemay, J. Kong, and C. Dekker, Nature {\bf 432}, 371 (2004).

\bibitem{sapmaz}
S. Sapmaz, P. Jarillo-Herrero, Ya.M. Blanter, C. Dekker, and
H.S.J. van der Zant, Phys. Rev. Lett. {\bf 96}, 026801 (2006).

\bibitem{ruitenbeek}
R.H.M. Smit, Y. Noat, C. Untiedt, N.D. Lang, M.C. van Hemert, and
J.M. van Ruitenbeek, Nature {\bf 419}, 906 (2002);
D. Djukic, K.S. Thygesen, C. Untiedt, R.H.M. Smit, K.W. Jacobsen,
and J.M. van Ruitenbeek, Phys. Rev. B {\bf 71}, 161402(R) (2005).

\bibitem{golubov}
A.A. Golubov, M.Yu. Kupriyanov, and E. Il'ichev,
Rev. Mod. Phys. \textbf{76}, 411 (2004).

\bibitem{chauvin}
M. Chauvin, P. vom Stein, D. Esteve, C. Urbina, J.C. Cuevas,
and A. Levy Yeyati, Phys. Rev. Lett. {\bf 99}, 067008 (2007).

\bibitem{novotny}
T. Novotn{\' y}, A. Rossini, and K. Flensberg, Phys.
Rev. B \textbf{72}, 224502 (2005).

\bibitem{zazu1}
A. Zazunov, R. Egger, C. Mora, and T. Martin,
Phys. Rev. B \textbf{73}, 214501 (2006).

\bibitem{bcs-holstein1}
J. Sk\"oldberg, T. L\"ofwander, V.S. Shumeiko,
and M. Fogelstr\"om, Phys. Rev. Lett. \textbf{101}, 087002 (2008).

\bibitem{zazu3} A. Zazunov, A. Schulz, and R. Egger, Phys. Rev.
Lett. \textbf{102}, 047002 (2009); A. Schulz, A. Zazunov, and
R. Egger, Phys. Rev. B {\bf 79}, 184517 (2009).

\bibitem{weig}
E.M. Weig, R.H. Blick, T. Brandes, J. Kirschbaum, W. Wegscheider,
M. Bichler, and J.P. Kotthaus, Phys. Rev. Lett. {\bf 92}, 046804 (2004).

\bibitem{garcia}
D. Garcia-Sanchez, A. San Paulo, M.J. Esplandiu, F. Perez-Murano, 
L. Forr{\'o}, A. Aguasca, and A. Bachtold, Phys. Rev. Lett. {\bf 99},
085501 (2007).

\bibitem{huettel}
A.K. H\"uttel, B. Witkamp, M. Leijnse, M.R. Wegewijs, and H.S.J. van der
Zant, Phys. Rev. Lett. {\bf 102}, 225501 (2009).

\bibitem{adrian}
B. Lassagne, Y. Tarakanov, J. Kinaret, D. Garcia-Sanchez, and A. Bachtold,
Science {\bf 325}, 1107 (2009).

\bibitem{weiss}
U. Weiss, \textit{Quantum dissipative systems}, 3rd edition
(World Scientific, Singapore, 2007).

\bibitem{alq} A. Zazunov, V.S. Shumeiko, E.N. Bratus', J. Lantz, and G. Wendin,
Phys. Rev. Lett {\bf 90}, 087003 (2003); A. Zazunov, 
V.S. Shumeiko, G. Wendin, and E.N. Bratus', Phys. Rev. B {\bf 71}, 214505 (2005).

\bibitem{LY}
A. Mart{\'i}n-Rodero, A. Levy Yeyati, and F.J. Garc{\'i}a-Vidal,
Phys. Rev. B {\bf 53}, R8891 (1996).

\bibitem{schattka} A. Schattka, Diploma Thesis, Heinrich-Heine Universit\"at
D\"usseldorf (2008).

\bibitem{gambetta}
A. Gambetta, C. Manzoni, E. Menna, M. Meneghetti, G. Cerullo, G. Lanzani,
S. Tretiak, A. Piryatinski, A. Saxena,  R.L. Martin, and A.R. Bishop,
Nat. Phys. {\bf 2}, 515 (2006).

\bibitem{phononexp} G.D. Sanders, C.J. Stanton, J.H. Kim, K.J. Yee,
Y.S. Lim, E.H. H{\'a}roz, L.G. Booshehri, J. Kono, and R. Saito, 
Phys. Rev. B {\bf 79}, 205434 (2009).

\end{thebibliography}
\end{document}